\documentclass{iopart}
\usepackage{times}
\usepackage[OT1]{fontenc}
\usepackage[latin1]{inputenc}
\usepackage{graphicx}
\usepackage{amssymb}
 
\begin{document} 

\title{Raman Spectroscopy of Mott insulator states in optical lattices}

\author{P. Blair Blakie}

\address{Jack Dodd Centre for Photonics and Ultra-Cold Atoms,
Department of Physics, University of Otago, P.O. Box 56, Dunedin, 
New Zealand}

\date{\today}

\ead{bblakie@physics.otago.ac.nz}

\begin{abstract}
We propose and analyse a Raman spectroscopy technique for probing the properties
of quantum degenerate bosons in the ground band of an optical lattice.
Our formalism describes excitations to higher vibrational bands and is valid for deep lattices where a tight-binding approach
can be applied to the describe the initial state of the system. In sufficiently deep lattices, localized states in higher vibrational bands play an important role in the system response, and shifts in resonant frequency of excitation are sensitive to the number
of particles per site. We present numerical results of this formalism applied to the case of a uniform lattice deep in the Mott insulator regime.
\end{abstract}
%\maketitle

\section{Introduction}
The observation of the Mott-insulator state of a degenerate Bose gas
in an optical lattice \cite{Greiner2002a} has opened an exciting
new avenue for investigating strongly correlated condensed matter
systems. Ensuing experiments (e.g. \cite{Greiner2002b,Mandel2003a,Widera2004a})
have investigated a wide variety of phenomena making optical lattices
one of the leading systems for studying quantum atom optics.

In the initial experimental report, two key pieces of evidence were
given for the Mott insulator state: the loss of phase coherence, and
the appearance of a gap in the excitation spectrum. More recent experimental work has used Bragg spectroscopy \cite{Schori2004a,Stoferle2004a} to study the equilibrium properties of this system. While this type of probing reveals the appearance of a gap in the excitation spectrum, experimental applications of this technique operate well-beyond the linear response regime, making direct comparison with theory difficult. \footnote{The observable for Bragg spectroscopy in optical lattices is the energy transferred to the system. This observable is rather difficult to measure accurately, and to obtain a signal experiments necessarily add a large amount of energy, well-beyond the linear response limit.}

In this paper we propose and theoretically analyse a Raman spectroscopy scheme for probing Mott insulator states in the lattice system. The formalism we present is sufficiently general to include the excitation of atoms to higher vibrational bands where they should be easily discernible from the unscattered atoms in time-of-flight analysis. We apply this formalism to a uniform lattice and show how the correlated nature of the Mott insulator state can give rise to localized states in the excited bands  when the lattice is sufficiently deep, and that the resonant frequency of these localized states is sensitive to the local filling factor in the optical lattice.

\section{Formalism} The system of interest, which we will refer
to as system 1, is a degenerate collection of bosonic atoms populating
the lowest vibrational band of an optical lattice well-characterized
by the Bose-Hubbard Hamiltonian

\begin{equation}
\hat{H}_{1}=\epsilon_{0}\sum_{j}\hat{n}_{j}-J\sum_{\langle i,j\rangle}\hat{a}_{i}^{\dagger}\hat{a}_{j}+\frac{U_{11}\alpha_{w}}{2}\sum_{j}\hat{n}_{j}(\hat{n}_{j}-1),\label{eq:HBH}\end{equation}
where $\hat{a}_{j}$ is a bosonic operator that annihilates an atom
from the Wannier state $w_{j}(\mathbf{x})$ centered on lattice site
$j$, with $\hat{n}_{j}=\hat{a}_{j}^{\dagger}\hat{a}_{j}$ the respective
number operator. The quantity $J$, known as the tunneling matrix
element, characterizes the tunneling between lattice sites and is
determined from band structure calculations \cite{Jaksch1998a,Blakie2004a}.
Interactions between particles are described by the matrix element
$\alpha_{w}\equiv\int d^{3}\mathbf{x}\,|w_{j}(\mathbf{x})|^{4}$ and
the coefficient $U_{11}=4\pi a_{11}\hbar^{2}/m,$ where $a_{11}$
is the s-wave scattering length for collisions between atoms in internal
state $1$. We restrict our attention here to the translationally
invariant system (i.e. neglect the influence of external trapping
potentials in addition to the optical lattice), and the constant $\epsilon_{0}$
characterizes the mid-point energy of the ground band.

\begin{figure}[tb]
\begin{center}
\includegraphics[width=9cm]{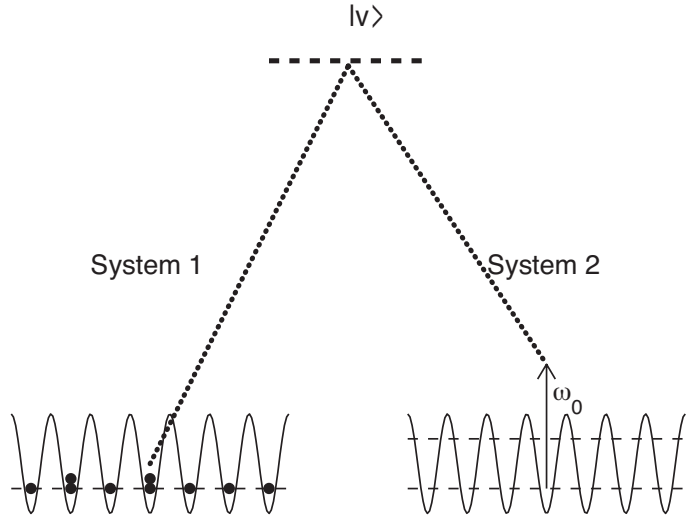}
\caption{\label{cap:exptdiag} Schematic diagram of the spectroscopic process. Atoms in the ground band of system 1 are transfered into the initially unoccupied system 2 by a two-photon Raman process through a virtual level $|v\rangle$. A residual energy difference $\hbar\omega_{0}$ and momentum kick $\hbar\mathbf{q}$ (not shown) are transfered to the atoms by each scattering event. 
}
\end{center}
\end{figure}

In this work, we consider a scheme for probing the properties of system
1, by an internal state changing Raman transition, of the form used
in Ref. \cite{Hagley1999a} to output couple an atom laser. A theory for this type of process in an optical lattices has also been proposed by Konabe \emph{et al.} \cite{Konabe2004a}\footnote{We note that the primary difference of our treatment to that of  Konabe \emph{et al.}, is that we include interactions between scattered and unscattered atoms, which are essential for the main results we present here.}. 
This is in contrast to a recent theory \cite{Menotti2003a,Roth2004a,Oosten2005a,Rey2005a,Batrouni2005a,Pupillo2006a} and experiments \cite{Schori2004a,Stoferle2004a} that have considered internal state preserving transitions to perform spectroscopy -- known as Bragg scpectroscopy \cite{Blakie2002a}. A schematic diagram of this coupling is shown in Fig. \ref{cap:exptdiag}.  Following the formalism in Ref. \cite{blakie2003a} this type
of coupling is described by an interaction term of the form\begin{equation}
\hat{V}=\frac{V_{p}}{2}\theta(t)\sum_{j}\int d^{3}\mathbf{x}\hat{\psi}_{2}^{\dagger}(\mathbf{x})\hat{a}_{j}w_{j}(\mathbf{x})e^{i(\mathbf{q}\cdot\mathbf{x}-\omega_{0}t)}+{\rm H.c},\label{eq:Hpert}\end{equation}
where $\hat{\psi}_{2}^{\dagger}$ is the field operator for bosonic
atoms in the second internal state, and the quantities $\hbar\mathbf{q}$
and $\hbar\omega_{0}$, specify the momentum and energy transfer of
the Raman process respectively. For simplicity we define $\hbar\omega_{0}$
to be the excess energy transferred over the internal state energy
difference. We take the amplitude of the Raman coupling to be of strength
$V_{p}$ beginning at $t=0,$ and of duration $T_{p}$. We will assume
that the internal states 1 and 2 are different hyperfine ground states
for which we can neglect any collisional spin evolution (see \cite{Widera2005a,Hall1998a}).
To arrange such a Raman probe experimentally, a pair of light fields (in addition to those used to create the lattice) with appropriately chosen polarizations to couple the states of interest will need to be applied to the atoms in the lattice. While such a probe could be focused down to address a few sites in a lattice, our theoretical development here is for the case where the Raman fields uniformly illuminate the system and have no spacial selectivity.

Our interest lies in the linear response regime, where only a small
portion of the atoms are scattered into internal state 2. The evolution
of these atoms in internal state 2, which we will refer to as system
2, is described by the Hamiltonian \begin{eqnarray}
\hat{H}_{2} & = & \int d^{3}\mathbf{x}\hat{\psi}_{2}^{\dagger}(\mathbf{x})(H_{{\rm sp}}+U_{12}\sum_{j}\hat{n}_{j}|w_{j}(\mathbf{x})|^{2})\hat{\psi}_{2}(\mathbf{x}),\label{eq:H2}\end{eqnarray}
 where 
 \begin{equation}H_{{\rm sp}}=p^{2}/2m+V_{{\rm ext}}(\mathbf{x}),\end{equation}
 is the single
particle Hamiltonian with $V_{{\rm ext}}(\mathbf{x})$ the external
potential (taken to be the same lattice potential experienced by atoms
in system 1), and $U_{12}=4\pi a_{12}\hbar^{2}/m$ characterizes the
interactions between particles in systems 1 and 2. Within the validity
regime of linear response it is permissible to neglect interactions
between atoms in system 2, as their density will remain low. 

System 2 is initially in the vacuum state, and the spectroscopic signal
to be measured is the number of atoms in system 2 after the Raman
pulse. This is a convenient experimental observable as detection techniques
can readily distinguish between atoms in different internal states
\cite{Hall1998a}. Additionally when the coupling produces atoms in
an excited vibrational band, the excited atoms can be differentiated
by their behavior upon expansion from the lattice \cite{Greiner2001a,Denschlag2002a}. These properties of the observable should enable the system response to be measured for small amounts of Raman excitation, allowing the system to be probed in the linear response regime.

The expression for the number of excited atoms, derived using linear
response theory, is\begin{equation}
\left\langle \hat{N}_{2}\right\rangle =\frac{\pi V_{p}^{2}T_{p}}{2\hbar^{2}}\int_{-\infty}^{+\infty}d\omega R(\mathbf{q},\omega)\,\frac{2\sin^{2}([\omega-\omega_{0}]T_{p}/2)}{\pi T_{p}(\omega-\omega_{0})^{2}},\label{eq:LinRespforN2}\end{equation}
where we have introduced the correlation function

\begin{eqnarray} R(\mathbf{q},\omega)=\frac{1}{2\pi}\int dt\int d^3\mathbf{x}\int d^3\mathbf{x}^\prime\,e^{i\mathbf{q}\cdot(\mathbf{x}^\prime-\mathbf{x})+i\omega t}\label{eq:RQWGen}\\ \times\sum_{ij}w_i^*(\mathbf{x})w_j(\mathbf{x}^\prime)\left\langle\hat{a}_i^\dagger(t)\hat{\psi}_2(\mathbf{x},t)\hat{\psi}_2^\dagger(\mathbf{x}^\prime,0)\hat{a}_j(0)\right\rangle_{\rm{Eq}}. \nonumber\end{eqnarray}In
this expression the time dependence of the operators is in an interaction
picture with respect to unperturbed Hamiltonians $\hat{H}_{1}+\hat{H}_{2}$,
and the expectation is taken on the initial equilibrium ensemble. 

We will now show how to evaluate the correlation function $R(\mathbf{q},\omega)$,
and how it relates to the properties of the atoms in system 1. We
take the initial density matrix to be $\rho=\rho_{1}\otimes\rho_{2}^{{\rm vac}}$,
where $\rho_{1}$ is the degenerate lattice state we are interested
in probing and $\rho_{2}^{{\rm vac}}$ is the initial vacuum state
for system 2. We can then show that without further approximation,
$R(\mathbf{q},\omega)$ can be written in the form

\begin{eqnarray}
R(\mathbf{q},\omega)&=&\frac{1}{2\pi}\int dt\int d^{3}\mathbf{x}\int d^{3}\mathbf{x}^{\prime}e^{i\mathbf{q}\cdot(\mathbf{x}^{\prime}-\mathbf{x})+i\omega t}\label{eq:RQ}\\&& \sum_{Q\, ij}\left\langle Q\left|\hat{a}_{i}^{\dagger}(t)\hat{a}_{j}(0)\rho_{1}\right|Q\right\rangle \left\langle {\rm 0}_{2}\left|\hat{\psi}_{j}^{Q}(\mathbf{x},t)\hat{\psi}_{2}^{\dagger}(\mathbf{x}^{\prime},0)\right|0_{2}\right\rangle w_{i}^{*}(\mathbf{x})w_{j}(\mathbf{x}^{\prime}),\nonumber\end{eqnarray}
where the variable $Q$ represents a trace carried
out over the number state basis for the operators $\{\hat{a}_{j}\}$,
i.e. $Q\leftrightarrow|\ldots,n_{l-1}^{Q},n_{l}^{Q},\ldots\rangle$,
and $|0_{2}\rangle$ represents the vacuum state of system 2. The
operator $\hat{\psi}_{j}^{Q}(\mathbf{x},t)$ is $\hat{\psi}_{2}(\mathbf{x})$
evaluated in an interaction picture with respect to the Hamiltonian\begin{equation}
\hat{H}_{j}^{Q}\equiv\int d^{3}\mathbf{x}\hat{\psi}_{2}^{\dagger}(\mathbf{x})(H_{{\rm sp}}+U_{12}\sum_{l}(n_{l}^{Q}-\delta_{lj})|w_{l}(\mathbf{x})|^{2})\hat{\psi}_{2}(\mathbf{x}).\label{eq:HjQ}\end{equation}
We note that in deriving Eq. (\ref{eq:RQ}) we have made use of the
result \begin{equation}
e^{i\hat{H}_{2}t/\hbar}\hat{\psi}_{2}(\mathbf{x})e^{-i\hat{H}_{2}t/\hbar}\hat{a}_{j}(0)|Q\rangle=\hat{a}_{j}(0)|Q\rangle\hat{\psi}_{j}^{Q}(\mathbf{x},t),\label{eq:HiQresult}\end{equation}
where the $\delta_{lj}$ term in Eq. (\ref{eq:HjQ}) arises from commuting
$\hat{a}_{j}$ with $\exp(i\hat{H}_{2}t/\hbar)$, and we have made
the replacement $\hat{n}_{l}\to n_{l}^{Q}$ as $|Q\rangle$ are number
states. 

We refer to $\hat{H}_{j}^{Q}$ as the $Q$-defect Hamiltonian for
system 2, as it arises from the removal of a system 1 atom from site
$j$ of the number state $Q$. As the system 1 atoms form an effective
potential for those in system 2, the removal of an atom at site $j$
due to the Raman excitation creates a potential hole (i.e. the $n_{l}^{Q}-\delta_{lj}$
term in Eq. (\ref{eq:HjQ})). This defect plays a key role in the
response spectrum as we will demonstrate later. Since $\hat{H}_{j}^{Q}$
is a quadratic Hamiltonian it can be diagonalized by numerical methods
to obtain its eigenvectors $\{\phi_{jm}^{Q}(\mathbf{x})\}$ and eigenvalues
$\{\hbar\omega_{jm}^{Q}\}$, i.e. $\hbar\omega_{jm}^{Q}\phi_{jm}^{Q}(\mathbf{x})=\hat{H}_{jQ}\phi_{jm}^{Q}(\mathbf{x}),$
where $m$ is the quantum number specifying the state and $j$ labels
the defect location. We note that in the limit $U_{12}\to0$, the
$\phi_{jm}^{Q}$ reduce to the Bloch states of $H_{{\rm sp}}$
and $m$ becomes the quasimomentum and band index. We obtain $\hat{\psi}_{j}^{Q}(\mathbf{x},t)=\sum_{m}\phi_{jm}^{Q}(\mathbf{x})\hat{b}_{jm}^{Q}e^{-i\omega_{jm}^{Q}t},$
where \textbf{$\hat{b}_{jm}^{Q}$} is a bosonic annihilation operator
and arrive at the expression
\begin{equation}
R(\mathbf{q},\omega)=\frac{1}{2\pi}\int dt\, e^{i\omega t}\sum_{Q}\sum_{ijm}c_{ij}^{Q}(t)A_{mjj}^{Q}A_{mij}^{Q*}e^{-i\omega_{jm}^{Q}t},\label{eq:RQ2}\end{equation}
where we have defined  
\begin{eqnarray}A_{mij}^{Q}&\equiv&\int d^{3}\mathbf{x}\,\phi_{jm}^{Q*}(\mathbf{x})e^{i\mathbf{q}\cdot\mathbf{x}}w_{i}(\mathbf{x}),\\
c_{ij}^{Q}(t)&\equiv&\langle Q|\hat{a}_{i}^{\dagger}(t)\hat{a}_{j}(0)\rho_{1}|Q\rangle.\end{eqnarray}

Eqs. (\ref{eq:RQ}) and (\ref{eq:RQ2}) represent the key results
of this work, and we now briefly comment on the physical process they
describe. Fundamentally, $R(\mathbf{q},\omega)$ characterizes the
excitation spectrum of atoms from system 1 into system 2. In the context
of ultra-cold gases, a result similar to our starting point {[}Eq.
(\ref{eq:LinRespforN2}){]} has been given by Luxat \emph{et al.}
\cite{Luxat2002a} for the case of a harmonically trapped Bose gas
(also see \cite{Choi2000a,Girardeau2001a}). However their treatment
neglects interactions between atoms in different hyperfine states
and assumes the single particle correlation functions for the atoms
in systems 1 and 2 are independent. The extension of this theory to the optical lattice has been provided by Konabe \emph{et al.} \cite{Konabe2004a}.
For the current experiments with Rubidium atoms in a deep
optical lattice these approximations are not tenable. In particular,
as we noted previously, the lattice site from which the atom is excited
acts as the localizing defect and accounting for correlations between
the systems (as we have done in Eq. (\ref{eq:RQ})) is essential. 

We note that our formalism [i.e. Eqs. (\ref{eq:RQ}) and (\ref{eq:RQ2})] is quite general. As long as the dominant number states of the many-body state are known, e.g. through exact diagonalization or Matrix Product Decomposition techniques (e.g. esee \cite{Clark2004a}), then the Raman response can be determined. In the following sections we consider the application of our formalism for two special cases. The results we present in the next section are calculated for 1D systems for numerical convenience, though our interest is in the regime where the scattering between particles is three-dimensional and well-described by a contact interaction.

\section{Single Site Limit} 
The limiting case of
a single tightly confining harmonic well of frequency $\omega_{{\rm ho}}$
is a useful approximation to a deep lattice in the regime where tunneling
between sites can be neglected. This limit shows the main physical
features of Raman spectroscopy and provides a useful approximation
to the full solution. Assuming that interaction shifts are small compared
to the oscillator energy $\hbar\omega_{{\rm ho}}$, we may approximate
the modes of the system as being harmonic oscillator eigenstates $\{\varphi_{m}(\mathbf{x})\}$,
with respective energies $\{\epsilon_{m}=\hbar\omega_{{\rm ho}}(m+1/2)\}$.
For the single site case our formalism maps according to
\begin{eqnarray}
 w_{j}(\mathbf{x})&\to&\varphi_{0}(\mathbf{x}),\\
\phi_{jm}^{Q}(\mathbf{x})&\to&\varphi_{m}(\mathbf{x}),\\
A_{mij}^{Q}&\to& A_{m}\equiv\int d^{3}\mathbf{x}\,\varphi_{m}^{*}(\mathbf{x})e^{i\mathbf{q}\cdot\mathbf{x}}\varphi_{0}(\mathbf{x}),\\
 \hbar\omega_{jm}^{Q}\to\hbar\omega_{m}^{n}&=&\epsilon_{m}+U_{12}\alpha_{0m}(n-1)\end{eqnarray}
where $\alpha_{0m}\equiv\int d^{3}\mathbf{x}\,|\varphi_{0}(\mathbf{x})|^{2}|\varphi_{m}(\mathbf{x})|^{2}$.
Since the Hamiltonian for system 1 reduces to $\hat{H}_{1}\to\epsilon_{0}\hat{n}+U_{11}\alpha_{w}\hat{n}(\hat{n}-1)/2$
in this limit, we have replaced $Q$ by $n$ (i.e. the single site many-body number states are just single mode number states $|n\rangle$) and evaluated the correlation
function as 
\begin{equation}c_{ij}^{Q}(t)\to c_{ij}^{n}(t)=n\exp(i[\epsilon_{0}+U_{11}\alpha_{w}(n-1)]t/\hbar).\end{equation}
Using Eq. (\ref{eq:RQ2}) with $\rho_{1}=\sum_{nn^{\prime}}|n\rangle\rho_{nn^{\prime}}\langle n^{\prime}|$,
we obtain\begin{eqnarray}
R(\mathbf{q},\omega)&=&\sum_{n}\rho_{nn}\, n|A_{m}|^{2}\delta(\omega_{{\rm res}}(n)-\omega),\label{eq:RqwSS} \\
\hbar\omega_{{\rm res}}(n)&=&\epsilon_{m}-\epsilon_{0}+[U_{12}\alpha_{0m}-U_{11}\alpha_{w}](n-1).\end{eqnarray}
We have also assumed that the probe only couples to a particular excited
state $m$, with the other states sufficiently far detuned that their
contribution is negligible. 

Eq. (\ref{eq:RqwSS}) shows that the response of the system is proportional
to $\rho_{nn}$, and most notably, if $[U_{12}\alpha_{0m}-U_{11}\alpha_{w}]\ne0$,
then the response frequency is linearly dependent on the value of
$n$ (i.e. the number of atoms at the site). In this regime the Raman
spectrum reveals the number distribution at the site. For the case
of $^{87}$Rb (the atom of primary interest in bosonic optical lattice
experiments) the interactions between the relevant hyperfine states
are approximately degenerate (i.e. $U_{11}\approx U_{12}$) , so that
the difference $\alpha_{0m}-\alpha_{w}$ will be the primary factor
in determining the magnitude of the number dependent shift. As an
immediate consequence we note that for the case $U_{11}=U_{12}$ and
a Raman pulse coupling to ground vibrational state of system 2, then
the spectrum will be independent of $n$ (i.e. for $m=0$ we have
$\alpha_{0m}=\alpha_{w}$). Therefore to obtain a number dependent
spectral response with $^{87}$Rb will require scattering into excited
vibrational states of system 2.

\section{Uniform Mott Insulator}  We now consider
the more general case of a translationally invariant lattice with
$N_{s}$ sites and periodic boundary conditions. We assume that the number of atoms in the system is commensurate with the number of lattice sites and
system is deep in the Mott insulating regime, where $U_{11}\alpha_{w}\gg J$. In this regime
the many-body ground state is well approximated as $|Q_{n}\rangle=|\ldots,n,n,\ldots\rangle$
(i.e a definite number of atoms at each site). In calculating the
response of the system to the Raman probe according to Eq. (\ref{eq:RQ}),
the summation over $Q$ reduces to this single state. The next order
correction to the translationally Mott state is particle hole states
\cite{Rey2005a}, which contribute to the ground state with an amplitude
$J/U_{11}\alpha_{w}\ll1$, and can be neglected. 

To obtain an approximation for the temporal correlation function of
system 1 in state $|Q_{n}\rangle$ we ignore the tunneling term in
$\hat{H}_{1}$. This approximation amounts to neglecting particle
tunneling between sites in system 1 over the time scale of the Raman
probe, and should be a good approximation in deep lattices. In this
limit the correlation function is 
\begin{equation}c_{ij}^{Q_{n}}(t)=n\,\exp(i[\epsilon_{0}+U_{11}\alpha_{w}(n-1)]t/\hbar)\delta_{ij},
\end{equation}
and making use of the translational invariance in evaluating Eq. (\ref{eq:RQ2})
we obtain
\begin{eqnarray}
R(\mathbf{q},\omega)&=&\sum_{m}n\, N_{s}|A_{mll}^{Q_{n}}|^{2}\delta(\omega-\omega_{{\rm res}}^{Q_{n}m}),\label{eq:RqLatt}\\
 \omega_{{\rm res}}^{Q_{n}m}&=&\omega_{lm}^{Q_{n}}-[\epsilon_{0}+U_{11}\alpha_{w}(n-1)]/\hbar.\end{eqnarray}
Note that due to translational invariance the precise value of $l$
used in Eq. (\ref{eq:RqLatt}) is unimportant. %
\begin{figure}[tb]
\begin{center}
\includegraphics[width=12cm]{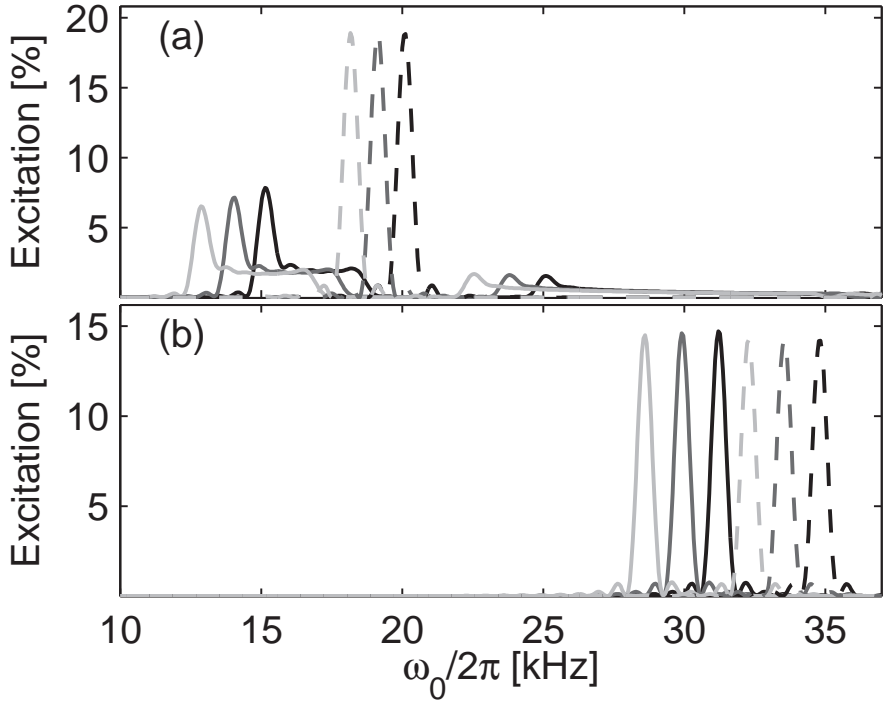}
\caption{\label{cap:SpecComp} The percentage of atoms excited after Raman
excitation in a 1D lattice of depth (a) $V_{D}=10E_{R}$ (b) $V_{D}=30E_{R}$.
In each plot the spectrum of Mott insulator states with filling factors
of $n=1$ (dark line), $n=2$ (medium line) and $n=3$ (light line)
are shown. The respective single site results for $m=1$ are given
as dashed lines. Parameters are $q=\pi/\lambda$, $T_{p}=1.5$ms,
$V_{p}=0.05E_{R}$, $\lambda=850$nm, $N_{s}=51$, transverse confinement
taken to be harmonic with $f_{\perp}\approx37.6$kHz, and $a_{11}=a_{12}=5.29$nm. }
\end{center}
\end{figure}
 We now calculate the response spectrum $\left\langle \hat{N}_{2}\right\rangle $
using Eqs. (\ref{eq:LinRespforN2}) and (\ref{eq:RqLatt}) for a translationally
invariant 1D lattice. We take the lattice potential to be $V_{{\rm ext}}(x)=V_{D}\cos^{2}(x/\lambda),$
arising from counter-propagating lasers fields of wavelength $\lambda$,
and we specify the lattice depth $V_{D}$ in units of $E_{R}=h^{2}/2m\lambda^{2}$.
The calculation is performed for typical $^{87}$Rb parameters (e.g.
see Ref. \cite{Greiner2002a}) to demonstrate the practicality of
our probing scheme. We will not consider the influence of $\mathbf{q}$
on the spectrum in this paper%
\footnote{Generally large $q$ favors coupling to higher bands, though certain
choices of $\mathbf{q}$ can suppress coupling efficiency, e.g for
$\mathbf{q}=\mathbf{0}$ the response from the first band is zero
by symmetry.%
}, and for simplicity we have taken $\mathbf{q}$ as being identical
to a reciprocal lattice for our calculations. 

The results we have obtained are shown in Fig. \ref{cap:SpecComp}.
For the case of $V_{D}=10E_{R}$ (Fig. \ref{cap:SpecComp}(a)) the
Raman response is shown over a frequency range that includes resonant
coupling to the first two excited bands. The superimposed graphs show
the spectra for Mott insulating states of various filling factors
and clearly exhibit frequency shifts proportional to $n$. Additionally,
we notice that in the first excited band ($12$kHz - $18$kHz), the
dominant response peak occurs at the low end of the spectral feature
(e.g. the peak at $\sim15.5$kHz for $n=1$), adjacent to a broad
base. This is feature is also seen in the 2nd excited band ($24$kHz
- $36$kHz), however the peaked feature is much less dominant relative
to the broad base. The peak originates from an excited band state
that is partially localized above the defect site. This localization
leads to a strong coupling matrix element $A_{mll}^{Q_{n}}$, and
as this state resides at the defect, its energy is lower than the
other states in the excited band. We refer to this state as the defect
state, which is clearly identified as the most strongly excited feature
at the bottom each spectral band. Since the localization is not perfect, the
other states in the excited band have appreciable amplitude at the
defect site. These states give rise to the broad though more weakly
excited, band of states above the resonant peak. In the 2nd excited
band, the same features are seen, however as the effective tunneling
in this band is much higher, and the defect state is much less localized. 

In Fig. \ref{cap:SpecComp}(b) the response spectrum is shown for
a lattice of depth $V_{D}=30E_{R}$. At this depth the response from
the first excited band states have shifted up to $30$kHz, and the
second excited band has moved out of the frequency range considered.
In contrast to the spectrum in Fig. \ref{cap:SpecComp}(a), we only
notice the resonant peak due to the defect state, without a discernible
broad base of band states. This arises because at this depth the tunneling
rate in the excited band is sufficiently small that the defect state
a becomes completely localized. The other states in the excited band
are necessarily orthogonal to the defect state and so have vanishing
coupling matrix elements $A_{mll}^{Q_{0}}$. 

For comparison in Figs. \ref{cap:SpecComp}(a) and (b) we also show
the single site predictions for the spectrum calculated using expression
(\ref{eq:RqwSS}) with $\omega_{{\rm ho}}$ chosen to match the effective
trap frequency at the lattice site minima. The frequency location
of the response spectra compares badly with the full lattice solution,
arising primarily from the inadequacy of the harmonic approximation
for accurately predicting band structure. The location of the spectra
can be easily corrected for by calculating the term $\epsilon_{m}-\epsilon_{0}$
in the expression for $\omega_{{\rm res}}(n)$ using the non-interacting
band structure result. However, the single site approximation captures
many of the salient features of the full lattice solution, such as
the magnitude of the $n$-dependent shift in the the response spectrum.
In the strongly-localized defect limit (Fig. \ref{cap:SpecComp}(b)),
the single site approximation quantitatively predicts the response
amplitude as the role of the band states can be neglected.%

\begin{figure}[tb]
\begin{center}
\includegraphics[width=12cm]{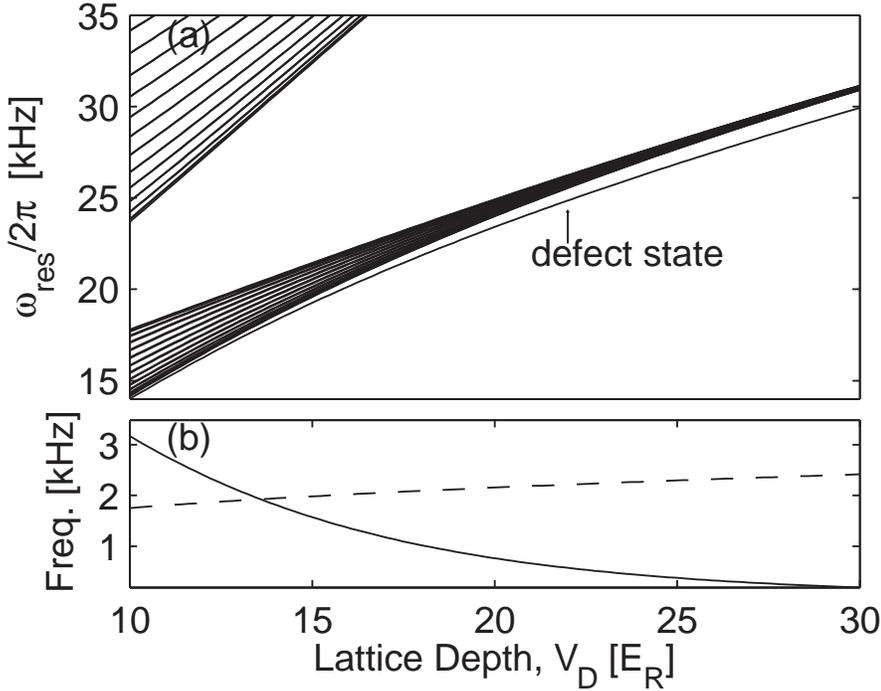}
\caption{\label{cap:SpecDefect} (a) Spectrum of excited states versus lattice
depth. (b) The bandwidth of the first excited band (solid) and the
on-site interaction (dashed) versus lattice depth. Other parameters
as in Fig. \ref{cap:SpecComp}.}
\end{center}
\end{figure} 

To quantify the emergence of the defect state we examine the spectrum
of $\omega_{{\rm res}}^{Q_{n}m}$ as a function of $V_{D}$ in Fig.
\ref{cap:SpecDefect}(a). For $V_{D}\gtrsim13.5E_{R}$ a single defect
state is observed to drop below the first excited band. The condition
for the emergence of a strongly localized defect state is that the
energy reduction from localization above the defect in system 1 is
large compared to the effective inter-site tunneling in the excited
band. To verify this criterion we compare the excited bandwidth (characterizing
the excited band tunneling rate) and energy reduction at the defect,
approximated by $U_{11}\alpha_{w}$. These two energy scales are seen
to cross at $V_{D}\approx13.5E_{R}$. Finally, we note the validity
condition for linear response treatment of Raman spectroscopy. This
requires the number of atoms excited to remain small compared to the
total number of atoms in the system. In terms of the Raman parameters
this condition is given as $nV_{p}^{2}T_{p}^{2}/4\hbar^{2}\ll1$,
where $n$ is the average number of atoms per site. 

We briefly comment on the relationship of the 1D calculations presented here to an equivalent system in a 3D lattice. The primary difference in applying our formalism to a fully three-dimensional system is that Eq. (\ref{eq:HjQ})  will need to be solved in 3D.  For the case of coupling to excited bands, the tunneling between lattice sites can occur in all directions, but  is dominated along the direction in which the vibrational excitation has occurred. This direction will be parallel to the direction of the momentum transfer in the Raman coupling, which we take to be parallel to a lattice vector. The tunneling in the orthogonal directions will be given by the ground band tunneling rate which is typically much smaller. This suggests that the additional shifts in a fully  3D lattice will be of order the ground state tunneling rate and will thus contribute small corrections to the results presented here. A full study in 3D will be the subject of a future investigation.

\section{Conclusions and outlook}
In this work we have proposed and analysed a Raman spectroscopy technique for probing the properties of quantum degenerate bosons in the ground band of an optical lattice.
We have observed that for sufficiently deep lattices, localized states in higher vibrational bands play an important role in the system response, and shifts in resonant frequency of excitation are sensitive to the number of particles per site. 
While our main study has considered the case of a perfect Mott insulator
in a translationally invariant lattice, our results suggest that in
the limit of strongly localized defect states the response of the
system is well-described by the single site result, and thus only
depends on the local number distribution at each lattice site. The
Raman spectrum may therefore be a useful method for measuring the relative portion
of system 1 at sites with filling factor $n$ atoms.
We speculate that in the strongly localized limit, the homogeneity of the lattice is rather unimportant, with the existence of the defect states arises from the large difference in the effective potential energy between the site where the particle is excited and the neighbouring site. This suggests that many of the predictions we have made here, and in particular the results of the 1-site model, should qualitatively apply to inhomogeneous lattices, such as the combined harmonic and optical lattice potentials made in experiments.  In this case, as well as in the superfluid limit (where significant number fluctuations exist), significant corrections may arise from resonances between neighbouring sites that would allow the defect to be localized over several sites. However, it seems reasonable to expect that these resonances would contribute to the broad background in the Raman spectrum, and the sharp features from Mott-insulating regions (if present) would be clearly visible. Characterizing the role of superfluid fluctuations and the external confining potential will be the subject of  future work.

\ack PBB would like to acknowledge valuable
discussions with Trey Porto, Crispin Gardiner, and thanks the University
of Otago for supporting this research.

\section*{References}
\bibliographystyle{unsrt}
\bibliography{raman}

\end{document}